\documentclass[fleqn,usenatbib]{mnras}
\usepackage{graphicx}
\usepackage[section]{placeins}
\usepackage{yfonts}

\usepackage{mathptmx}
\usepackage{times} 

\usepackage[T1]{fontenc}
\usepackage{amsmath}
\usepackage{amssymb}
\usepackage{tikz}
\usepackage{xcolor}

\begin{document}

\title[Orbit injection of planet-crossing asteroids]{Orbit injection of planet-crossing asteroids}
\author[F. Namouni]{F. Namouni$^{1}$\thanks{E-mail:namouni@oca.eu} \\
$^{1}$Universit\'e C\^ote d'Azur, CNRS, Observatoire de la C\^ote d'Azur, CS 34229, 06304 Nice, France}

\date{Accepted 2023 November 15. Received 2023 November 03; in original form 2023 August 25}
\pubyear{0000}
\maketitle

\begin{abstract}
Solar system Centaurs originate in transneptunian space from where planet orbit crossing events inject their orbits inside the giant planets' domain. Here, we examine this injection process in the three-body problem  by studying the  orbital evolution of transneptunian asteroids located at Neptune's collision singularity as a function of the Tisserand invariant, $T$.  Two injection modes are found, one for $T>0.1$, or equivalently prograde inclinations far from the planet, where unstable motion dominates injection, and another for $T\leq 0.1$, or equivalently polar and retrograde inclinations far from the planet,  where stable motion dominates injection. The injection modes are independent of the initial semi-major axis and the dynamical time at the collision singularity. The simulations uncovered a region in the polar corridor where the dynamical time exceeds the solar system's age suggesting the possibility of  long-lived primordial {polar transneptunian reservoirs that supply Centaurs} to the giant planets' domain. 

\end{abstract}

\begin{keywords}
celestial mechanics--comets: general--Kuiper belt: general--minor planets, asteroids: general. \\[-5mm]
\end{keywords}

\section{Introduction}
Understanding precisely how the dynamical transfer of transneptunian objects into the giant planets' domain  takes place is key  to identifying the origin of the small body populations in the outer solar system \citep{Fernandez16, Fraser22, KaibVolk22}. {

In this regard, progress was recently spurred by the long-term statistical simulations of high-inclination Centaurs that demonstrated that their likely past orbits clustured  perpendicular to the solar system's invariable plane in a region that extends beyond Neptune's orbit termed `the polar corridor' \citep{NamouniMorais18b, NamouniMorais20b, NamouniMorais20c}. 

The search  for the polar corridor's dynamical origin and its role in the transfer of  transneptunian objects (TNOs) onto  high inclination Centaur orbits over Gyr timescales was the object of our earlier work  (\cite{Namouni22}, hereafter Paper I).  By studying analytically and numerically the long-term dynamics of a planet-crossing asteroid in the three-body problem, we demonstrated that  despite the highly chaotic nature of the asteroid-planet interactions, the Tisserand invariant is conserved and that for moderate to high inclinations, this invariance makes the asteroid follow the Tisserand inclination pathway given as: 
\begin{equation}
 I(a,T,a_p)=\arccos \left(\left[T-\frac{a_p}{a}\right]{\left[4\left(2-\frac{a_p}{a}\right)\right]^{-\frac{1}{2}}}\right), \label{TissIncP}
\end{equation}
where  $a$ and $a_p$ are the semi-major axes of the asteroid's and the planet's orbits, $T$ is the Tisserand invariant:
\begin{equation}
T=\frac{a_p}{a}+2 \left[\frac{a(1-e^2)}{a_p}\right]^\frac{1}{2} \cos I, \label{TissRel}
\end{equation}
and $e$ and $I$ are the asteroid orbit's eccentricity and inclination. It was also demonstrated that the inclination pathway (\ref{TissIncP}) is independent of the initial perihelion and aphelion.

The motion of planet-crossing asteroids follows the Tisserand inclination pathway whether time flows forwards or backwards on the timescale of Gyrs. The pathway is followed strictly when the gravitational kicks imparted to the asteroid's orbit induce significant change to its semi-major axis. When the orbit's semi-major axis is not strongly perturbed the asteroid's orbit wanders away from the inclination pathway at constant Tisserand invariant because of the build-up of secular perturbations, only to return to the Tisserand inclination pathway. 

Paper I demonstrated orbit injection of incoming asteroids, from outside to inside the planet's orbit, onto polar or retrograde inclination orbits occurs only for $-1<T<2$. Translated in terms of  the Tisserand pathway's inclination far from the planet:  
\begin{equation}
I_\infty(T)=\arccos (T/\sqrt{8}),\label{Iinfty}
\end{equation}
the previous statement  becomes: only TNOs with inclinations $45^\circ<I_\infty<110^\circ$ may enter inside the planet's orbit and assume inclinations $ I\geq 90^\circ$.   TNOs with $I_\infty<45^\circ$ may be injected inside Neptune's orbit but can only assume prograde orbits whereas those with $I_\infty>110^\circ$ may not be injected at all.  The long-term statistical simulations of high-inclination Centaurs (\citep{NamouniMorais20b} hereafter Paper II) utilised $2\times 10^7$ clones to follow the Centaurs' evolution back to $-4.5$\,Gyr from initial inclinations in the range $[62^\circ:173^\circ]$.  They yielded average inclinations for the Centaurs' end states away from Neptune in the range $[60^\circ:90^\circ]$ with a standard deviation $\sim 10^\circ$ indicating that the polar corridor corresponds approximately to the analytically determined range of $-1<T<2$. 

Paper I established a novel dynamical behaviour of planet-crossing asteroids. The decreasing semi-major axis of an incoming asteroid, whose planet encounters move it towards the planet, cannot be smaller than a minimum value that depends on its Tisserand invariant. The asteroid's motion is reflected back at this minimum value, termed  `the reflection semi-major axis', ${a_{\rm refl}}$, meaning that the asteroid's  semi-major axis increases after approaching  ${a_{\rm refl}}$ which is given as:
\begin{equation}
{a_p}{a^{-1}_{\rm refl}(T)} = T - 2 + 2 ( 3- T)^\frac{1}{2}.\label{aX}
\end{equation}
The smallest reflection semi-major axis occurs for $T=2$ at ${a_{\rm refl}}=a_p/2$ whereas for $T<-1$, ${a_{\rm refl}}>a_p$ and reflection occurs outside the planet's orbit thus precluding orbit injection. The  validity of the Tisserand inclination pathway and the reflection semi-major axis holds for  $T\lesssim 2.7$ or $I_\infty >17^\circ$ even though the Tisserand invariant is always conserved regardless of its value. This limitation is related  to the effect of strong mean motion resonances at low inclinations (see Paper I).

The present work is the continuation of the analysis in Paper I. We examine precisely how  Neptune-crossing asteroids are injected from outside to inside its orbit.  }Studying the dynamical pathways numerically in Paper I utilised the cloning of an asteroid's orbit with a perihelion (aphelion) equal or smaller (larger) that Neptune's semi-major axis. Whereas this approach is adequate to establish the conservation of the Tisserand invariant and  study the dynamics of the Tisserand inclination pathways, it does not provide a comprehensive analysis of the injection process over a large set of diverse initial conditions.  In this work, we follow a different approach by studying the dynamics at Neptune's collision singularity.  

In the three-body problem, the collision singularity is defined by the orbital intersection of the Neptune-crossing asteroid with the planet. For an initial Tisserand invariant and an asteroid semi-major axis, a diverse range of eccentricities, inclinations and orbital angles can be  generated to cover the collision singularity, and followed over timespans larger than the solar system's age. As the Tisserand invariant is conserved, the diverse asteroid orbits that cover the collision singularity should in principle follow the same Tisserand inclination pathway and allow us to examine the orbit injection process as a function of the Tisserand invariant. 

In Section 2, the initial conditions setup at the collision singularity is explained. In Section 3, we examine the dynamics at the collision singularity for stable and unstable orbits and find that the Tisserand invariant conservation at the collision singularity ensures that the statistical ensemble's initially diverse orbits follow the same Tisserand inclination pathway. The dynamical time at the collision singularity is measured and found to exceed the solar system's age for certain negative Tisserand invariant values  or equivalently retrograde inclinations at infinity.
The evolution of the minimum semi-major axis at the collision singularity is examined and reveals a semi-major axis-independent two-mode injection process, one at retrograde inclinations at infinity where stable motion dominates injection and another at prograde inclinations at infinity where unstable motion dominates injection. We also examine the role of mean motion resonance capture in the collision singularity and find that it cannot explain overall stability and the two-mode injection process. In Section 4, we measure the extent of the collision singularity's stable region in the polar corridor and identify the possible location of  Centaur supplying small body reservoirs beyond Neptune's orbit. In Section 5, we sum up our findings and point out how the presence of  long lived primordial polar TNO reservoirs was already indicated in the earlier numerical simulations with all giant planets in Paper II.

\section{Initial conditions at the collision singularity}
The collision singularity is defined by the intersection of the  three-dimensional orbit of the planet-crossing asteroid with the planet's circular orbit.  Intersection may occur at the ascending or descending nodes and is expressed by equating the asteroid's radial distance at its nodes with the planet's semi-major axis as follows:
\begin{equation}
r=\frac{a(1-e^2)}{1\pm\cos\omega}=a_p, \label{nodes}
\end{equation}
where $a$ and $a_p$ are the asteroid's and the planet's semi-major axes, and $e$ and $\omega$ are the astereoid's eccentricity and argument of perihelion. The $\pm$ signs correspond to the ascending or descending nodes respectively with $0^\circ\leq\omega\leq180^\circ$.

 For an initial Tisserand invariant $T$ and an initial semi-major axis $a$, we generate random initial conditions that place the asteroid on an intersecting trajectory as follows.  The asteroid's mean longitude $\lambda$, longitude of ascending node $\Omega$, and $\omega$ are drawn from uniform distributions in the intervals $[-180^\circ:180^\circ]$ for the first two and  $[0^\circ:180^\circ]$ for the third angle. Eccentricity is obtained from the nodal distance equation (\ref{nodes}) to ensure orbit intersection as:
\begin{equation}
 e_{\rm coll}=\frac{\left[4a(a-a_p)+a_p^2\cos^2\omega\right]^\frac{1}{2}\pm a_p\cos\omega}{2a}, \label{eccinitcond}
\end{equation}
where the $-$ ($+$) sign corresponds to a crossing at the ascending (descending) node.  The orbit's inclination is obtained from the Tisserand invariant (\ref{TissRel}) using the initial semi-major axis and the collision eccentricity (\ref{eccinitcond}) as follows:
\begin{equation}
I=\arccos\left((T-a_p/a){\left[4a(1-e_{\rm coll}^2)/a_p\right]^{-\frac{1}{2}}}\right). \label{incinitcond}
\end{equation}

The minimum eccentricity of the collision singularity ensemble generated by this method is that of an encounter at perihelion $e_{\rm min}=1-a_p/a$. The maximum eccentricity is determined by the Tisserand invariant  as the amplitude of the argument in the brackets of (\ref{incinitcond}) should be $\leq 1$. This leads to the maximum eccentricity encountered in Paper I and given by:
\begin{equation}
e_{\rm max}=\left(1-\frac{a_p}{4a}\left[T-\frac{a_p}{a}\right]^2\right)^\frac{1}{2}. \label{emax}
\end{equation}
The existence of a maximum eccentricity reduces the argument of perihelion range depending on the value of $T$. Figure 1 shows the collision eccentricity for encounters at the ascending and descending nodes  for an initial semi-major axis $200$\,au that intersects Neptune's orbit with $a_p=30.1$\,au. Maximal eccentricity is shown for  $T=1.5,\ 2.5$ and $2.15$. The latter value is the limit {below} which there is no gap in the argument of perihelion range. It is given generally as $T=2+a_pa^{-1}.$ The $\omega$-gap is shown in the Figure as the shaded region for $T=2.5$. 

The inclination range is related to the eccentricity's by the Tisserand invariant. At minimum eccentricity, inclination is that of the Tisserand inclination pathway (\ref{TissIncP}) whereas at maximum eccentricity, inclination is $0^\circ$ for prograde orbits and $180^\circ$ for retrograde orbits. 
\begin{table}

\centering
\caption{Initial eccentricity and inclination ranges $[e_{\rm min}:e_{\rm max}]$ and $[I_{\rm min}:I_{\rm max}]$. $e_{\rm min}$ is the eccentricity for perihelion crossing, $e_{\rm max}$ is given by (\ref{emax}), $I_{\rm min}$ and $I_{\rm max}$ are given by (\ref{TissIncP}) and either of $0^\circ$ (prograde motion) or $180^\circ$ (retrograde motion). $I_\infty$ is given by (\ref{Iinfty}). }
\label{table:1}
\begin{tabular}{ccccc}
\hline 
\multicolumn{5}{c}{$a_0=100$\,au, $e_{\rm min}=$0.699}\\
$T_0$  & $e_{\rm max}$ &  $I_{\rm min}(^\circ)$ &  $I_{\rm max}(^\circ)$ &  $I_\infty(^\circ)$ \\
\hline
$-0.7$    &     0.961        &     113   &    180        &      104\\
$-0.1 $    &         0.993    &           99  &     180     &           92\\
  1.2      &        0.969     &          0      &    70        &         65\\
  2.7       &      0.753      &         0         & 23           &      17\\
\hline
\multicolumn{5}{c}{$a_0=200$\,au, $e_{\rm min}=$0.849}\\
$-0.8 $    &    0.983&             104   &   180 &               106\\
$-0.3$     &  0.996      &         99     & 180       &            96\\
1.0         &    0.986         &        0   &    72           &      70\\
2.5        &      0.980           &      0 &      30             &     28\\
 \hline
 \multicolumn{5}{c}{$a_0=73$\,au, $e_{\rm min}=$0.588}\\
$-0.5$  &          0.956    &          111&      180 &           100\\
0        &           0.991       &         99  &   180    &          90\\
1.4      &          0.948        &         0    &   67      &          60\\
2.8       &        0.642          &       0      & 19        &      8\\
\hline
\end{tabular}
\end{table}

\section{Collision singularity dynamics}
The Tisserand invariant conservation was established in Paper I by generating clones of a number of planet-crossing asteroid orbits. For one such asteroid with a Tisserand invariant $T_0$,  the cloned orbits differ slightly from the nominal orbit and the deviations of their Tisserand invariants were measured with respect to the nominal orbit.  In the present work, the initial conditions for an initial $T_0$ are  different as there is no nominal orbit. By letting the initial angle variables sample their maximum possible range subject to the limits set by the Tisserand invariant $T_0$, the generated statistical ensemble with an initial semi-major axis $a_0$ is made up of different asteroids with diverse eccentricities and inclinations and not clones of the same asteroid. 

The equations of motion of the circular restricted three-body problem with Neptune at $a_p=30.1$\,au and  collision singularity asteroids at $a_0=73$\,au, $100$\,au and $200$\,au were integrated forward in time using the Bulirsch-Stoer {algorithm} with an error tolerance of $10^{-11}$. In this simulation, the Tisserand invariant  takes values in the range $-1.2\leq T_0\leq 2.8$ with a 0.1-step. { The choice of this range is motivated by two particular values of the Tisserand invariant. One is $T_0=-1$ the upper inclination boundary of the polar corridor below which asteroids are reflected back outside the planet's orbit $a_{\rm refl}>a_p$ and injection is not possible. The other particular value is $T_0=2.7$ above which mean motion resonances induce a significant departure of  the inclination pathway from $I_\infty$ even though the Tisserand invariant is conserved (see Paper I). }

For each $T_0$, 1000 orbits were generated. {The eccentricity and inclination ranges for each semi-major axis $a_0$ produced by the method in Section 2 are given in Table 1.} The integration timespan of 5.4\,Gyr is larger than the solar system's age because the simulation unveiled a region in parameter space where the dynamical time exceeds 4.5\,Gyr.  As in Paper I, orbital evolution stops when the asteroid is ejected, collides with the Sun or Neptune, or reaches an outer semi-major axis boundary at $10^4$\,au.  For each asteroid orbit, the minimum semi-major axis is monitored.

\subsection{Tisserand inclination pathways}

{Figure (2)} shows snapshots of the orbit distribution of the collision singularity ensembles with $a_0=100$\,au for four values of the Tisserand invariant $T_0=2.7,\ 1.2,\ -0.1$ and $-0.7$. {These examples illustrate the dynamics corresponding to $I_\infty$ from small and moderate prograde to polar and retrograde motion.}  Stable orbits are shown in green and unstable orbits in red.\footnote{Stable orbits live the full duration of the simulation ($\geq 4.5$\,Gyr). Unstable orbits either collide with the sun or the planet or are ejected from the system. Their lifetimes range from a few 1,000 years to $<4.5$\,Gyr.} Also shown are the Tisserand inclination pathway (\ref{TissIncP}), the maximum eccentricity curve (\ref{emax}), and the reflection semi-major axis (\ref{aX}). 

The top row panels show the eccentricity and inclination as a function of semi-major axis at 4.5\,Gyr for stable orbits and at the last sampling epoch before instability for unstable orbits. The conservation of the Tisserand invariant makes the diverse ensemble of collision singularity asteroids follow the unique Tisserand 
inclination pathway given by their initial $T_0$ as predicted by the analytical theory of Paper I. Their eccentricity and inclination dispersions are explained by the Von-Zeipel-Lidov-Kozai secular potential \citep{Lidov62,Kozai62,VonZeipel10} as evidenced by maximum eccentricity curve of each $T_0$ (see Paper I for an analytical discussion of this effect). Orbits are predictably reflected at $a_{\rm refl}(T_0)$.  These conclusions are valid for stable and unstable orbits as the latter are simply former stable orbits for which an ultimate gravitational kick knocked them out of their stable motion by ejecting them from the system  or forcing them to collide with the sun or Neptune.  Instability is stronger as $T_0$ tends to 3. 

The location $a_0=100$\,au is close to the 1:6 outer resonance ($a_{\rm res}=99$\,au) where some asteroids are captured  mostly at large inclinations ($T_0=1.2$, $-0.1$ and $-0.7$) as mean motion resonance capture is particularly efficient at large inclinations \citep{NamouniMorais15,NamouniMorais17}.  Some collision singularity asteroids with retrograde inclinations become unstable while at resonance. Resonance capture is examined in Section 3.4 as a function of the Tisserand invariant. 

The panels in the second row from the top show the eccentricity and inclination at minimum semi-major axis, $a_{\rm min}$.  The asteroids follow their Tisserand inclination pathway  in ($a_{\rm min}$,$I$) space  just as they do in ($a$,$I$) space and are bounded by $e_{\rm max}$ in ($a_{\rm min}$,$e$) space. Stable and unstable asteroids alike fill the possible area in ($a_{\rm min}$,$I$) and ($a_{\rm min}$,$e$) showing that not all asteroids reach $a_{\rm refl}(T_0)$ or are injected inside the planet's orbit.

 The third row panels show the epoch of minimum semi-major axis, $T_{\rm a_{min}}$. This epoch is bounded from below by a power-like function $\log (\tau_{\rm min}/\tau_0)=(a_0-a)^\alpha$ where $a$ and $a_0$ are expressed in au, $\alpha\sim 0.5$ and $\tau_0\sim 10^3$ to $10^5$\,yr with decreasing $T_0$. The travel time to minimum semi-major axis is longer for retrograde orbits {because planet encounters generate weaker perturbations} as they occur at larger speeds in briefer times than prograde orbits.   Asteroids captured in resonance have some of the largest $T_{\rm a_{min}}$. Resonant asteroid statistics are given in Section 3.4.

The bottom row panels show the initial conditions' distribution in the mean longitude, argument of perihelion plane for stable and unstable orbits. Asteroids that achieved orbit injection are shown with blue squares. An $\omega$-gap is present for $T_0=2.7$. There are no preferred regions or initial condition clustering with respect to stability or orbit injection. 

In Section 3.3, we examine the dynamical time in the collision singularity and show that there is a Tisserand invariant range where the dynamical time exceeds the solar system's age for $a_0=100$\,au. 
We therefore set up two further high resolution simulations, one with $a_0=200$\,au where the dynamical time nowhere exceeds 4.5\,Gyr and another with $a_0=73$\,au so that the initial location is not close to a major mean motion resonance like 100\,au is to the 1:6 resonance. {The scope of these two simulations is to find out if the dynamics at different semi-major axes differ qualitatively from that of $a_0=100$\,au because of stability and proximity to resonance. }

The simulation results with $a_0=200$\,au are shown in Fig. 3 for $T_0=2.5$, 1, 0.3 and $-0.8$. They confirm that the collision singularity asteroids follow the Tisserand inclination pathways in the $(a,I)$ and $(a_{\rm min},I)$ planes for stable and unstable orbits. Their eccentricities are bounded by the maximal eccentricity (\ref{emax}).  The location $a_0=200$\,au is close to the 1:17 mean motion resonance ($a_{\rm res}=199$\,au) where similar to the previous simulation some asteroids are captured  mostly at large inclinations ($T_0=1$ and $-0.3$) whereas some collision singularity asteroids become unstable while at resonance. The dynamics at $a_0=200$\,au is qualitatively similar to that of $a_0=100$\,au.

The value $a_0=73$\,au was chosen instead of the rounded off $70$\,au  to avoid the 2:7 resonance. Instead  $a_0=73$\,au lies within the 9:34 resonance of order 25.  The simulation results are shown in Fig. 4  for $T_0=2.8$,  1.4, 0 and $-0.5$. $T_0=2.8$  is below the validity limit of the Tisserand inclination pathways $T=2.7$ (see Paper I). {It is shown to illustrate as in Paper I that reflection still occur even if $T_0>2.7$ but the actual  the reflection semi-major axis become slightly smaller than (\ref{aX}) (for details see Paper I).} The results in Fig. 4 are qualitatively similar to those of the previous two semi-major axes. Inclination pathways and maximum eccentricity curves are followed precisely for $T_0\leq 2.7$ and there is no preferred orbit geometry that maximises orbit injection (bottom panels in Fig. 2, 3 and 4).  Some asteroids are captured in the 9:34 resonance (see Section 3.4). 

We may conclude that regardless of semi-major axis and proximity to mean motion resonance, the collision singularity asteroids are found to follow the Tisserand inclination pathways\footnote{The Tisserand inclination pathway is followed in the sense explained in the Introduction either strictly if semi-major axis changes are large or as an inclination boundary if secular perturbations build up. See examples in Fig. 10.} and the maximal eccentricity curves associated with their Tisserand invariant. Their orbits fill the available parameter space in the ($a_{\rm min},I)$- and  ($a_{\rm min},e)$-planes. There are no preferred orbital angles that maximize  orbit injection. 

\subsection{Tisserand invariant conservation}
Deviation from the initial value of the Tisserand invariant is measured at 4.5\,Gyr and shown in Fig.~5. The solid lines show the deviation of the mean Tisserand invariant, $\langle T\rangle$ from its initial value $T_0$ for stable and unstable orbits as a function of $T_0$ for $a_0=73$\,au, 100\,au and 200\,au. The deviation $|\langle T\rangle-T_0|$ {is of} order a few $10^{-3}$ and is smaller for unstable orbits than stable orbits. The reason is twofold. First, unstable orbits are more numerous as can be seen  in Fig. 6 where the ratio of the number of stable orbits $N_{\rm s}$ to that of unstable orbits $N_{\rm u}$ is shown as a function of $T_0$ ({and $I_\infty$}). Second, accumulation of numerical error is larger for stable orbits that survived up to 4.5\,Gyr whereas unstable orbits live over smaller timespans from $\sim 10^3$\,yr to less than $4.5$\,Gyr. This explains why $|\langle T\rangle-T_0|$ increases as $T_0$ tends to $-1$ where stable orbits become more numerous. 

The standard deviation $\sigma_T$ of $T$ at 4.5\,Gyr is shown in Fig. ~5 with dashed lines for stable and unstable motion. It is of order a few $10^{-3}$ across all values of $T_0$.  This deviation amplitude as well as that of   $|\langle T\rangle -T_0|$   do not affect quantitatively the dynamics of the Tisserand inclination pathways as can be seen in Figs. 2, 3 and 4.

\subsection{Dynamical times}
For an initial semi-major axis $a_0$ and Tisserand invariant $T_0$, we define the singularity collision dynamical time, $\tau_{\rm d}(a_0,T_0)$,  as the median lifetime of the statistical ensemble generated by the method in Section 2. This definition is inspired by the long term evolution of the ensemble's diverse orbits that cover the available parameter space defined by $a_0$ and $T_0$ (Figs. 2, 3, and 4) much like the clones of single asteroid orbits in Paper I. Figure 7 shows the dynamical time as a function of $T_0$  ({and $I_\infty$}) for $a_0=73$\,au, 100\,au, and 200\,au. 

Regardless of semi-major axis, the dynamical time has a distinct shape with three peaks: (i) a prominent retrograde-inclination one with $-0.85\leq T_0\leq -0.6$ equivalent to an inclination at infinity $107^\circ\geq I_\infty \geq 102^\circ$, (ii) an intermediate prograde-inclination peak with $0.7\leq T_0\leq 1$ ($76^\circ\geq I_\infty \geq 69^\circ$), and (iii) a shallow prograde-inclination peak with $2.1\leq T_0\leq 2.3$ ($42^\circ\geq I_\infty \geq 36^\circ$). 

{The dynamical time is of order }$0.1$-$0.4$\,Gyr near the shallow peak. It increases to 1.5-2\,Gyr at the intermediate peak. The retrograde inclination peak occurs at $\tau_{\rm d}=4.11$\,Gyr and $T_0=-0.9$ for $a_0=200$\,au.  It exceeds the age of the solar system for $a_0=73$\,au and $a_0=100$\,au with respectively $\tau_{\rm d}=5.37$\,Gyr at $T_0=-0.5$, and  $\tau_{\rm d}=5.20$\,Gyr at $T_0=-0.8$. Regardless of semi-major axis, the dynamical time drops sharply from a few Gyr to $\sim 1$\,Gyr for polar orbits ($T_0=0$).

The region where the dynamical time exceeds 4.5\,Gyr is $-1.1\leq T_0 \leq -0.3$  for $a_0=73$\,au and $-1\leq T_0 \leq -0.4$ for $a_0=100$\,au. This raises the possibility of long-lived { nearly polar TNOs } that cross Neptune's orbit. {Injection is examined in Section 3.5 and  the stable region's semi-major axis extent in Section 4. }

\subsection{Resonance capture}
The simulations show that resonance capture may occur for collision singularity asteroids.  It is possible that the large dynamical times found at retrograde inclinations that exceed  the solar system's age are the result of such temporary resonance capture as asteroid orbits are protected from the planet's destabilizing  perturbations. We examined quantitatively the simulations of the three semi-major axes and found that capture occurs mainly in the mean motion resonance nearest to the initial semi-major axis. These are  9:34 for $a_0=73$\,au, 1:6 for $a_0=100$\,au and 1:17 for $a_0=200$\,au. {Figure 8} shows for each initial semi-major axis, the ratio of the number of stable asteroid orbits in mean motion resonance with Neptune, $N_{\rm res}$,  to the number of stable asteroid orbits, $N_{\rm s}$, as a function of the initial Tisserand invariant $T_0$ at epochs 0.1\,Gyr, 1\,Gyr, 3\,Gyr and 4.5\,Gyr. 

Regardless of semi-major axis and resonance order (5, 16 and 25), the resonant asteroid ratio has three major peaks. For $a_0=73$\,au, the resonant ratio peaks  $(T_0,N_{\rm res}/N_{\rm s})$ are $(0,16\%)$, $(0.7,19\%)$ and  $(1.4,6\%)$. For $a_0=100$\,au the peaks are $(-0.1,30\%)$, $(0.7,34\%)$ and  $(1.4,36\%)$ and for $a_0=200$\,au, they are $(-0.3,20\%)$, $(0.5,28\%)$ and  $(1.5,6\%)$. The similar peak locations with respect to the Tisserand invariant  indicate the existence of preferred  inclinations for resonance capture within the polar corridor.  

Another property of the resonant ratio is how it conserves the same profile from epoch 0.1\,Gyr to 4.5\,Gyr for $a_0=100$\,au and $a_0=200$\,au and  from 1\,Gyr to 4.5\,Gyr for $a_0=73$\,au. The profile amplitudes increase with time confirming that resonances increase the stability of trapped collision singularity asteroids. The reason {why at} $a_0=73$\,au the resonant ratio profile of epoch 0.1\,Gyr is different at later epochs  {is that the initial semi-major axis lies} inside the 9:34 resonance region $[72.94,73.08]$\,au. Dynamical relaxation over $\sim 1$\,Gyr  shifts the main retrograde inclination peak from $T_0\sim-0.3$ to $T_0=0$. 

For low inclinations, it is the 1:6 resonance that is more efficient at resonance capture as seen on the sharp increase in the resonant ratio. Of the 1000 asteroids at $a_0=100$\,au and $T_0=2.8$,  4 asteroids survived at 4.5\,Gyr one of which in the 1:6 resonance. The 1:17 and 9:34 resonant ratios essentially vanish beyond $T_0\sim2$. 

The resonant ratios show that resonance capture is not the main reason for  the longevity of retrograde inclination collision singularity asteroids. In the $\geq 4.5$\,Gyr $T_0$-range of $a_0=73$\,au, $[-1.1:-0.3]$, the resonant ratio $N_{\rm res}/N_{\rm s}\sim5\%$ whereas for the equivalent range of $a_0=100$\,au, $[-1:-0.4]$,  $N_{\rm res}/N_{\rm s}\sim 20\%$. These estimates do not explain the longevity of the bulk of orbits with $\tau_{\rm d}\geq4.5$\,Gyr. 

\subsection{Orbit injection}
{Paper I showed that orbit injection occurred only if   $T>-1$. Injected TNOs from the polar corridor ($-1\leq T\leq 2$ or equivalently $45^\circ\leq I_\infty\leq 110^\circ$) may assume polar and retrograde orbits whereas those with  $T>2$ (or equivalently $I_\infty \leq 45^\circ$) may be injected  only onto prograde orbits. }

Fig.~9 shows the ratio, $N^{\rm i}_{\rm s}/N^{\rm i}_{\rm u}$, of the number of injected stable orbits present at epoch 4.5\,Gyr to the number of all injected unstable orbits from the simulation's start to 4.5\,Gyr {for the three semi-major axes $a_0=$73\,au, 100\,au and 200\,au}. The Tisserand invariant range starts from $T_0=-0.9$, because injection is not possible for $T_0\leq -1$ for stable and unstable orbits alike ($N^{\rm i}_{\rm s}=N^{\rm i}_{\rm u}=0$)  as $a_{\rm refl}>a_p$ (see Paper I). 

{Regardless of semi-major axis, there are two injection modes depending on the Tisserand invariant. For $T_0> 0.1$ ($I_\infty <88^\circ$), most injected asteroids have unstable orbits.  For $T_0\leq 0.1$ ($I_\infty\geq 88^\circ$), most injected orbits are stable. } This property is independent of the dynamical time being larger than the solar system's age. For instance, whereas for $a_0=200$\,au, $\tau_{\rm d}=3.67$\,Gyr and  $N_{\rm s}/ N_{\rm u}=0.8$ at $T_0=-0.6$, the injection ratio $N^{\rm i}_{\rm s}/N^{\rm i}_{\rm u}= 8$. Regardless of semi-major axis, the injection ratio may be approximated by  $\log N^{\rm i}_{\rm s}/ N^{\rm i}_{\rm u}=-T_0/0.61$ over a wide range of Tisserand invariants.  {It should also be noted that the two-injection modes are present within the polar corridor and that the unstable-injection mode for prograde motion is unchanged as the polar corridor's $T_0=2$ boundary is crossed.}

Combined with the $>4.5$\,Gyr dynamical times for negative $T_0$ at certain semi-major axes, the stable injection mode implies the possible existence of long-lived primordial {nearly polar } TNO reservoirs that supply high inclination Centaurs to the giant planet's domain. {The semi-major extent of this stable region is examined in the next Section.  We further illustrate the stable injection mode in  Figure 10 with two examples of 4.5\,Gyr-stable collision singularity TNOs with an initial semi-major axis $a_0=100$\,au and an initial Tisserand invariant $T_0=-0.5$ that are transferred onto retrograde  Centaur orbits.  In the left (right) two panels the TNO's  initial parameters are $e_0=0.7112$, $I_0=108^\circ$, $\omega_0=154^\circ$ ($e_0=0.8313$, $I_0=113^\circ$, $\omega_0=92^\circ$). The first TNO (panels a and b) is injected early on during its interaction with the planet in two Centaur phases. The first occurs in the interval  $[0.14:0.2]$\,Gyr  where its orbital parameters $e\sim 0.1$-$0.2$ and $I=150^\circ$. The second phase occurs in $[0.26:0.39]$\,Gyr with  $e\sim 0.05$-$0.65$ and $I=140^\circ$-$178^\circ$.  At 4.5\,Gyr the TNO's semimajor axis $a=375$\,au, the location of the 1:44 resonance with Neptune that the asteroid entered at epoch 4.12\,Gyr. The conservation of the Tisserand invariant ensures that the TNO follows the inclination pathway (\ref{TissIncP}) and its eccentricity is limited by the maximal eccentricity curve (\ref{emax}). 

After moving away from Neptune as far as 230\,au, the second TNO in panels (c, d) is injected onto a Centaur orbit in the interval $[3.97:4.11]$\,Gyr where its  mean orbital elements  are $e\sim 0.4$ and $I\sim 160^\circ$. At 4.5\,Gyr the TNO's semi-major axis $a=54$\,au. The TNO's evolution differs significantly from that of the previous one except in one aspect, the conservation of the Tisserand invariant makes them follow the same Tisserand inclination pathway (\ref{TissIncP})  and their eccentricity is always limited by (\ref{emax}).

}

\section{Stable collision singularity region}
We examine the possible extent of the long-lived region with negative Tisserand invariants in the Sun-Neptune-asteroid three-body problem by following the evolution of collision singularity ensembles within the grid $-1\leq T_0\leq 0$ with a 0.1-step and 40\,au$\leq a_0\leq$200\,au with a 10\,au-step. Each grid point ensemble includes $100$ asteroids\footnote{Except for $a_0=100$\,au and 200\,au whose ensembles include 1000 asteroids.}  generated by the method of Section 2 and integrated forward in time to 5\,Gyr because of the computational cost in the grid's stable regions. 

The dynamical time portrait as a function of the initial semi-major axis $a_0$ and the Tisserand invariant, $T_0$ is shown in Fig.~11. Color codes indicate the dynamical time, $\tau_{\rm d}$. The simulation timespan 5\,Gyr is also the maximal value of  $\tau_{\rm d}$. Grid points where $\tau_{\rm d}=5\,$Gyr may have larger dynamical times.  For our purposes of searching for possible 4.5\,Gyr-old TNO reservoirs that supply high inclination Centaurs to the giant planets' domain, the 5\,Gyr timespan  is sufficient. 

{The stable region ($\tau_{\rm d}\geq4.5$\,Gyr) can be seen to extend up to $a_0=170$\,au. Beyond this limit,   possible material present in the early solar system would have been depleted as $\tau_{\rm d}<4.5$\,Gyr. For each semi-major axis inside the stable region, there are islands of Tisserand invariants where the dynamical time  is much larger than the solar system's age.  }The region enclosed by  $-0.9\leq T_0\leq-0.5$ ($\leq I_\infty\leq $) and $80$\,au$\leq a\leq 140$\,au is a continuous domain where $\tau_{\rm d}\geq 5$\,Gyr. In this domain, the initial semi-major axes are close to the mean motion resonances 3:13 at 80\,au and  1:$n$ with $5\leq n\leq 10$ for $a_0= 90$\,au to 140\,au. In Section 3.4, it was shown that mean motion resonances do not explain the overall stability of collision singularity asteroids for $-1\leq T_0\leq 0$ as only a fraction of the long-lived asteroids are captured in resonance. 

{In order to identify the possible primordial reservoirs that inject TNOs onto Centaur orbits, we show in Figure 12  the ratio of the number of injected stable orbits at  4.5\,Gyr, $N^{\rm i}_{\rm s}$ to the initial asteroid number, $100$ (or $1000$), only for locations where  $\tau_{\rm d}\geq 4.5$\,Gyr and the injected number of stable orbits is finite ($N^{\rm i}_{\rm s}>0$).

Two regions are of particular interest. The first is nearest to the planet 40\,au$\leq a_0\leq$80\,au (blue and green squares) where injection ratios are largest 15$\%$ to 30$\%$. The second region is the 5\,Gyr continuous stable domain. Here the injection ratio peaks at 10$\%$ for $a_0=90$\,{au and} 100\,au  and $-0.9\leq T_0\leq -0.5$.  The finite injection ratios shown in Figure 12 indicate that all such locations are prime contenders for hosting long-lived primordial nearly polar TNO reservoirs that supply Centaurs to the solar system.  However, as we explain in the next section, the stable region nearest the planet is likely to loose its stability and large injection ability when  all giant planets are added to the Sun-Neptune-TNO three-body problem.}

\section{Summary}
In this work, we set out to study the injection dynamics of transneptunian asteroids that cross Neptune's orbit in the three-body problem. Injection in this framework means the process of lowering the value of the asteroid's semi-major axis so that it becomes smaller than the planet's. In order to model this process comprehensively, we examined the dynamics of Neptune's collision singularity at different transneptunian semi-major axes by using the conservation of the Tisserand invariant for planet-crossing asteroids established in Paper I to generate statistical ensembles at the collision singularity and follow their evolution. 

We examined the dynamics of stable and unstable motion at the collision singularity and uncovered a two-mode injection process regardless of the initial semi-major axis and the dynamical time. For polar and retrograde inclinations, stable motion dominates injection whereas for prograde motion unstable motion dominates. There are regions outside Neptune's orbit where the median lifetime of retrograde collision singularity asteroids exceeds the solar system's age. We also found that the collision singularity's dynamical time has a unique dependence on the Tisserand invariant regardless of semi-major axis. Similarly, capture in resonance of collision singularity asteroids is more efficient at preferred inclinations in the polar corridor regardless of semi-major axis. 

These results' validity holds in the three-body problem. The solar system's giant planets will affect the injection process and the collision singularity's stability. {This influence can be constructive or destructive.  Pluto's Neptune-crossing orbit provides an example of Jupiter's constructive influence on perihelion precession that helps  confine its latitudinal motion  \citep{Malhotra22}. 
In the case of the polar TNO reservoirs uncovered in this work, Paper II indicates that the presence of all giant planets modifies the dynamics at high inclination both constructively and destructively.

To understand this, we recall that Paper II aimed at searching for past stable orbits of 19 high-inclination Centaurs by integrating the equations of motion back in time to $-4.5$\,Gyr with all giant planets and the Galactic tide using $\sim 2 \times 10^7$ clones. }The idea behind that search was that since Centaurs were believed to have originated from the early planetesimal disc, their orbits are 4.5\,Gyr old.  Searching statistically for the region in parameter space such objects came from is possible in principle because of the time-reversibility of the gravitational equations of motion. {In fact, the conservation of the Tisserand invariant makes the inclination pathway taken by a planet-crossing asteroid unchanged whether its motion is followed forwards or backwards in time on Gyr timescales as shown later in Paper I. 
The statistical stable orbit search did not indicate that high inclination Centaur orbits returned to the planetesimal disk, nor was it inconclusive such as when past orbits  scattered all over parameter space.} Instead, high-inclination Centaur  clustered mainly in the polar corridor with small inclination dispersions on Neptune-crossing orbits (Fig. 2 and Fig. 4 of Paper II).

One of the search's unexpected results was that the Centaur semi-major-axis  probability density at $-4.5\,$Gyr has a prominent peak between 70\,au and 100\,au (Fig. 3 in Paper II) indicating the possible existence of a TNO reservoir where the three-body analysis in the present paper predicts.  {However, despite the simulation's high resolution of  $2\times 10^7$ clones, there was no accumulation in the region nearest to Neptune (40\,au$\leq a\leq$70\,au) where the three-body analysis of this paper found the largest injection rates. This  indicates that the presence of all giant planets likely destroys the stability in that region.} 

{The large inclination  of  the possible 4.5 Gyr-old TNO reservoirs poses the problem of their origin as at that early epoch, transneptunian space was devoid of solar system-born material orbiting the sun perpendicular to the planetesimal disk. Evidence from planet instability-based disk relaxation simulations  over 4.5\,Gyr indicate that high-inclination Centaurs cannot be produced in sufficient numbers from the early planetesimal disk, and that other sources are necessary to explain their presence \citep{kaib19,Nesvorny19}. The natural source is the material captured by the early solar system from the sun's birth cluster \citep{Fernandez00,Levison10,Brasser12b,Jilkova16,Hands19,kaib19}. The present work demonstrates, in the context of the three-body problem, the possible existence of pockets of dynamical stability where some primordial interstellar material is protected in the solar system and interacts with its planets as polar and retrograde Centaurs.  Understanding further collision singularity dynamics with the giant planets  and how they influence the stable region,  found in this paper using the three-body problem, will help narrow the extent of the stability region and ascertain the population size of the possible Centaur-producing polar TNO reservoirs in the solar system.

}

\section*{Acknowledgments}
I thank the reviewer for useful comments. The numerical simulations were done at the M\'esocentre SIGAMM hosted at the Observatoire de la C\^ote d'Azur. 
 \bibliographystyle{mnras}

\vspace*{-3mm}

\section*{Data availability}
The data underlying this article will be shared on reasonable request to the  author.

\vspace*{-5mm}

\bibliography{ms}
\newpage
\newpage
\begin{figure}
{ \begin{center}
\includegraphics[width=70mm]{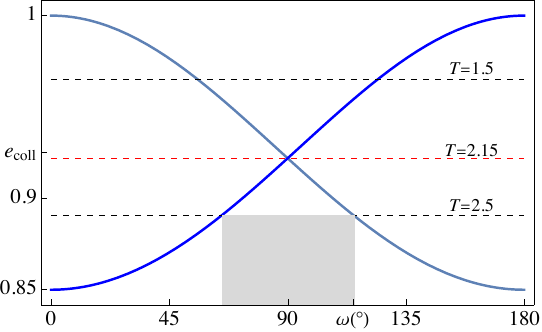}\label{f1}
\end{center}
}
\begin{center}
\caption{{Collision eccentricity (\ref{eccinitcond}) as a function of argument of perihelion for ascending (solid blue) and descending {(solid dark blue)} nodes with an asteroid at $a=200$\,au. Maximal eccentricity  (\ref{emax}) is shown as a horizontal dashed line for three values of the Tisserand invariant. The shaded region is the $\omega$-gap below $T=2.5$.}}
\end{center}\label{f1}
\end{figure}

\begin{figure*}
{ 
\hspace*{-42mm}\includegraphics[width=250mm]{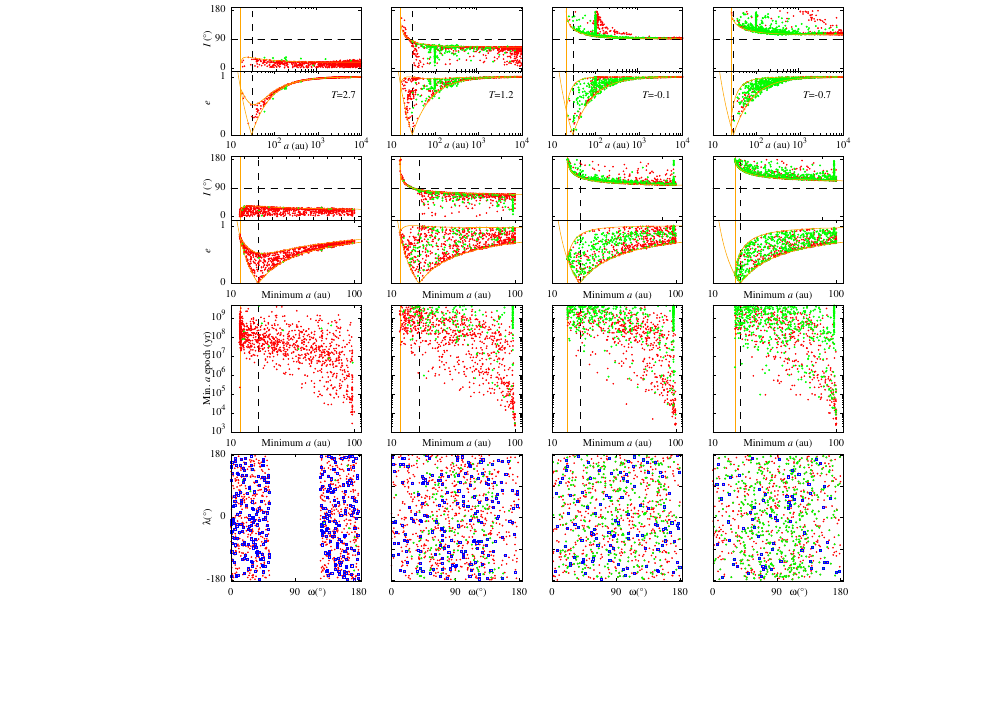}\hspace{10mm} \label{f2}\vspace*{-30mm}
}
\begin{center}
\caption{Distribution of collision singularity ensemble orbits at $4.5\,$Gyr for different initial Tisserand invariants $T_0$ and the initial semimajor axis $a_0=100$ au. {\it First row from top:} Distributions in the $(a,I)$ and $(a,e)$ planes of stable (green) and unstable (red) orbits. {The latter are shown at the last sampling epoch before instability.} The inclination pathway (\ref{TissIncP}) is the solid curve in the inclination panels. In the eccentricity panels, the top curve is the maximum eccentricity dispersion (\ref{emax}) and the bottom ones are the perihelion and aphelion collision conditions  $a(1\pm e)=a_p$. The solid and dashed vertical lines indicate the reflection semimajor axis $a_{\rm refl}$ (\ref{aX}) and the planet's position respectively. {\it Second row:} Distributions in the $(a_{\rm min},I)$ and $(a_{\rm min},e)$ planes. {\it Third row:} Epoch at minimum semi-major axis as a function of $a_{\rm min}$. {\it Bottom row:} Initial conditions in the mean longitude-argument of perihelion plane. The blue squares indicate injection was achieved.}
\end{center}\label{f2}
\end{figure*}

\begin{figure*}
{ 
\hspace*{-42mm}\includegraphics[width=250mm]{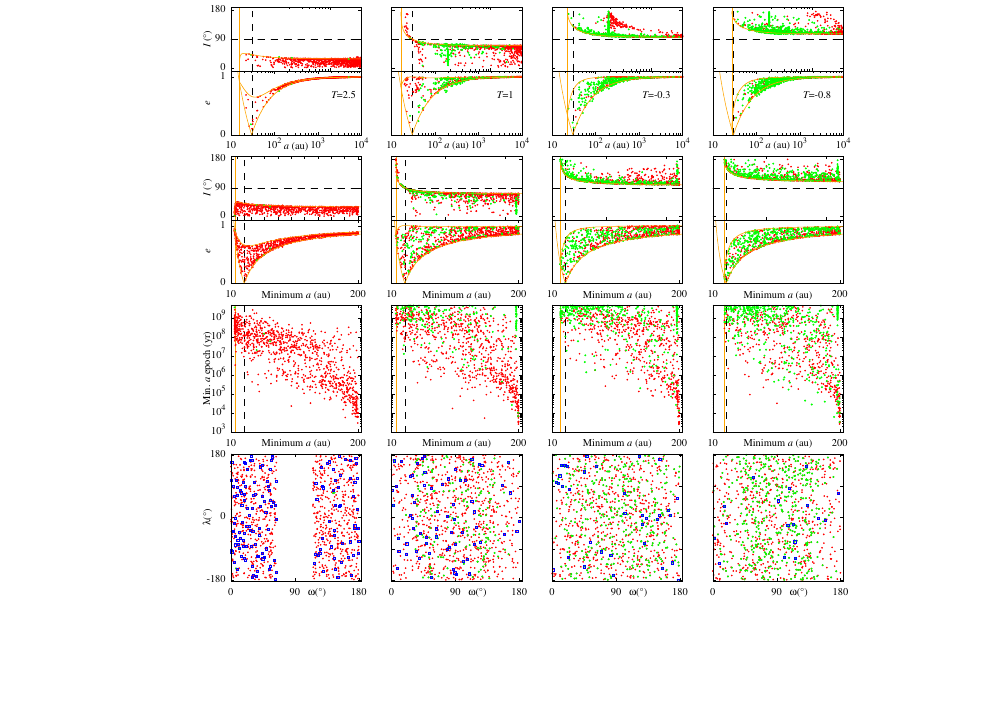}\hspace{10mm} \label{f3}\vspace*{-30mm}
}
\begin{center}
\caption{Same as Fig.2 with $a_0=200$\,au.}
\end{center}\label{f3}
\end{figure*}

\begin{figure*}
{ 
\hspace*{-42mm}\includegraphics[width=250mm]{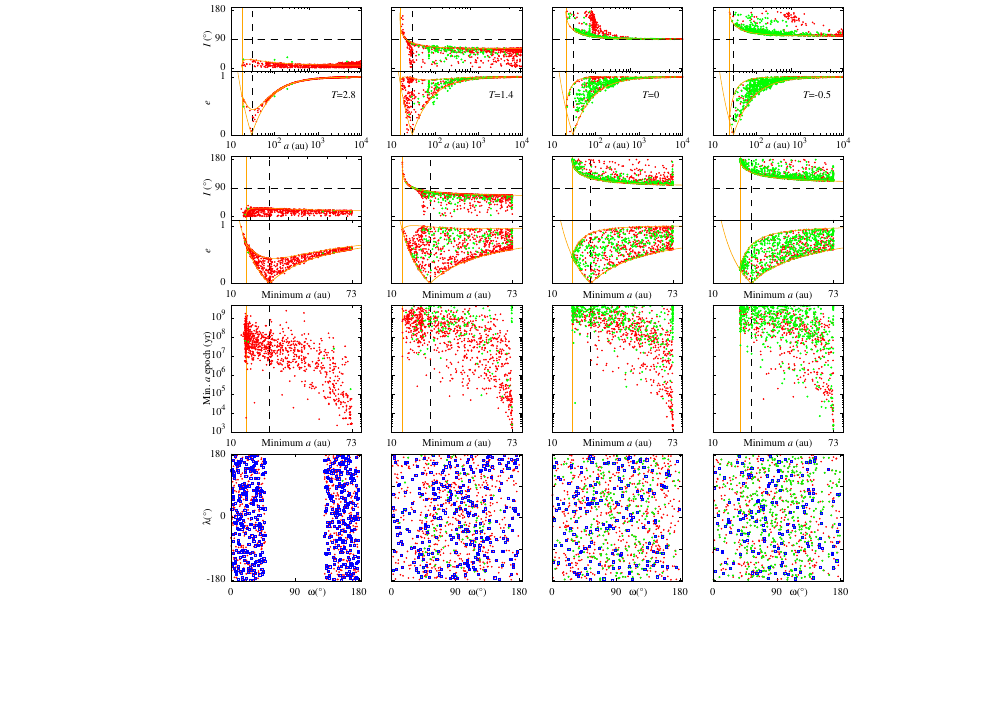}\hspace{10mm} \label{f4}\vspace*{-30mm}
}
\begin{center}
\caption{Same as Fig.2 with $a_0=73$\,au.}
\end{center}\label{f4}
\end{figure*}

\begin{figure}
{ 
\hspace*{-5mm}\includegraphics[width=90mm]{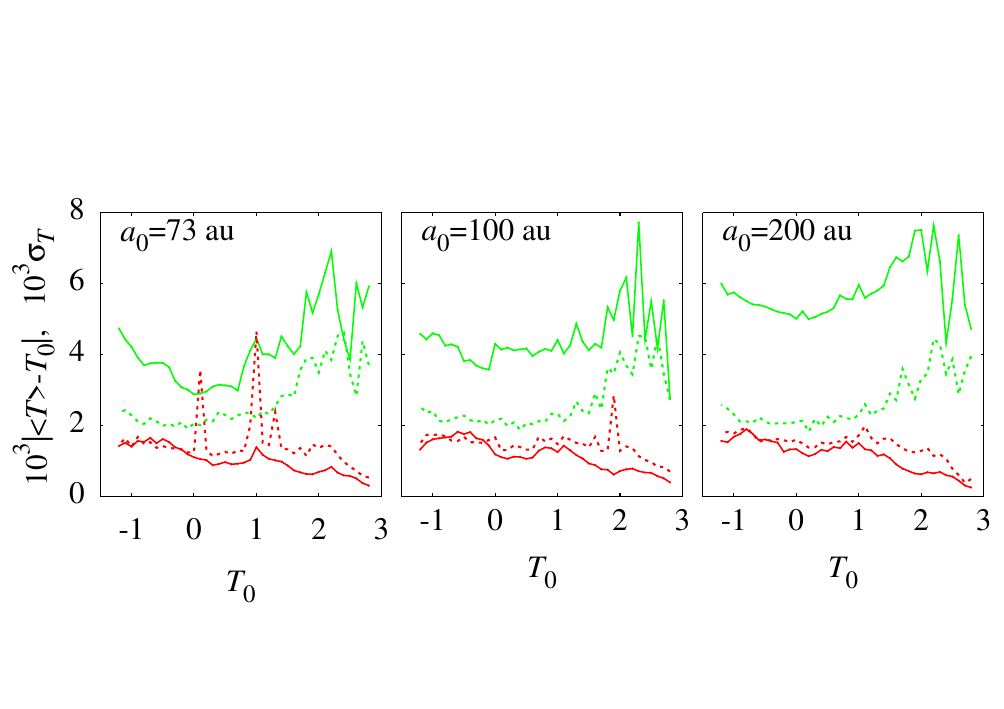} \label{f5}\vspace*{-3mm}\vspace*{-10mm}
}
\begin{center}
\caption{Conservation of the Tisserand invariant at $4.5$\,Gyr. Solid lines indicate the mean value of $T$ with respect to $T_0$ and dashed lines indicate the Tisserand invariant's standard deviation $\sigma_T$. Green (red) indicates stable (unstable) orbits. }
\end{center}\label{f5}
\end{figure}

\begin{figure}
{ 
\includegraphics[width=80mm]{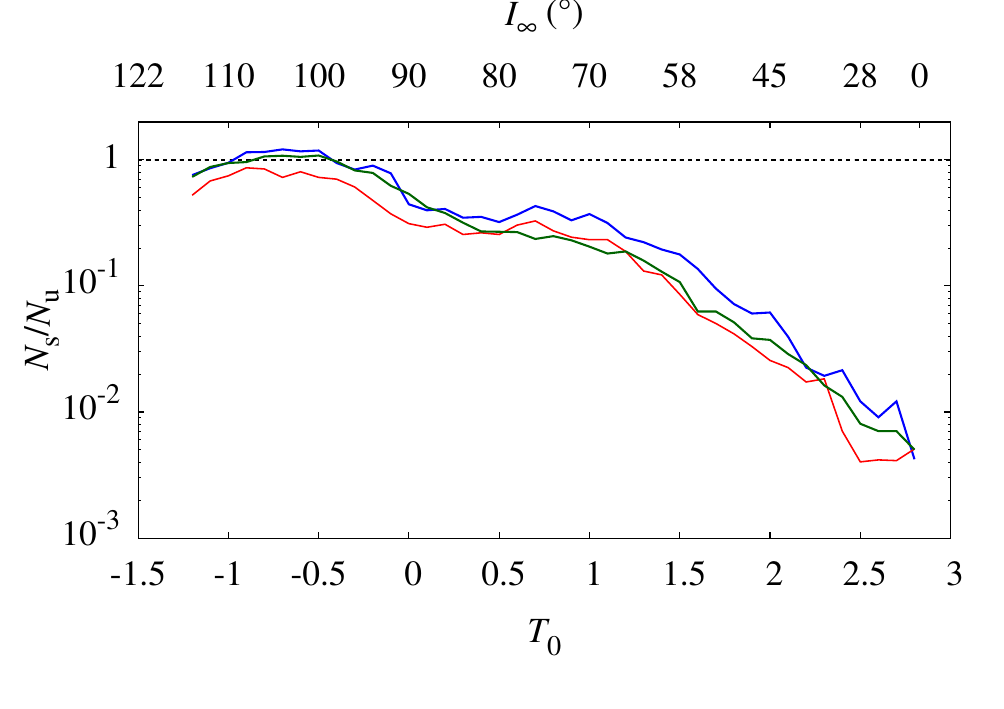} \label{f6}\vspace*{-3mm}
}
\begin{center}
\caption{Ratio of the number of stable orbits present at $4.5$\,Gyr, $N_{\rm s}$, to the number of all unstable orbits from the simulation's start up to $4.5$\,Gyr, $N_{\rm u}$, as a function of the initial Tisserand invariant (and $I_\infty$) . The solid curves correspond respectively to $a_0=200$\,au (red), $100$\,au (blue) and $73$\,au (green).}
\end{center}\label{f6}
\end{figure}

\begin{figure}
{ 
\includegraphics[width=85mm]{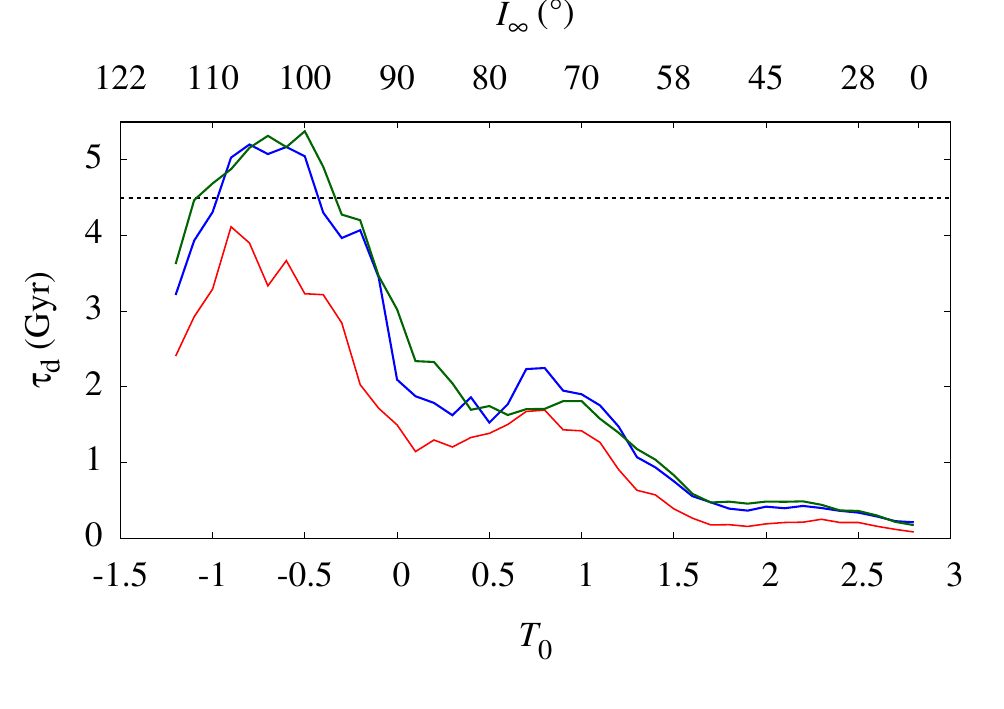} \label{f7}\vspace*{-3mm}
}
\begin{center}
\caption{Dynamical time $\tau_d$ as a function of the initial Tisserand invariant (and $I_\infty$)  for $a_0=200$\,au (red), $100$\,au (blue) and $73$\,au (green).}
\end{center}\label{f7}
\end{figure}

\begin{figure}
{ 
\hspace*{-2mm}\includegraphics[width=90mm]{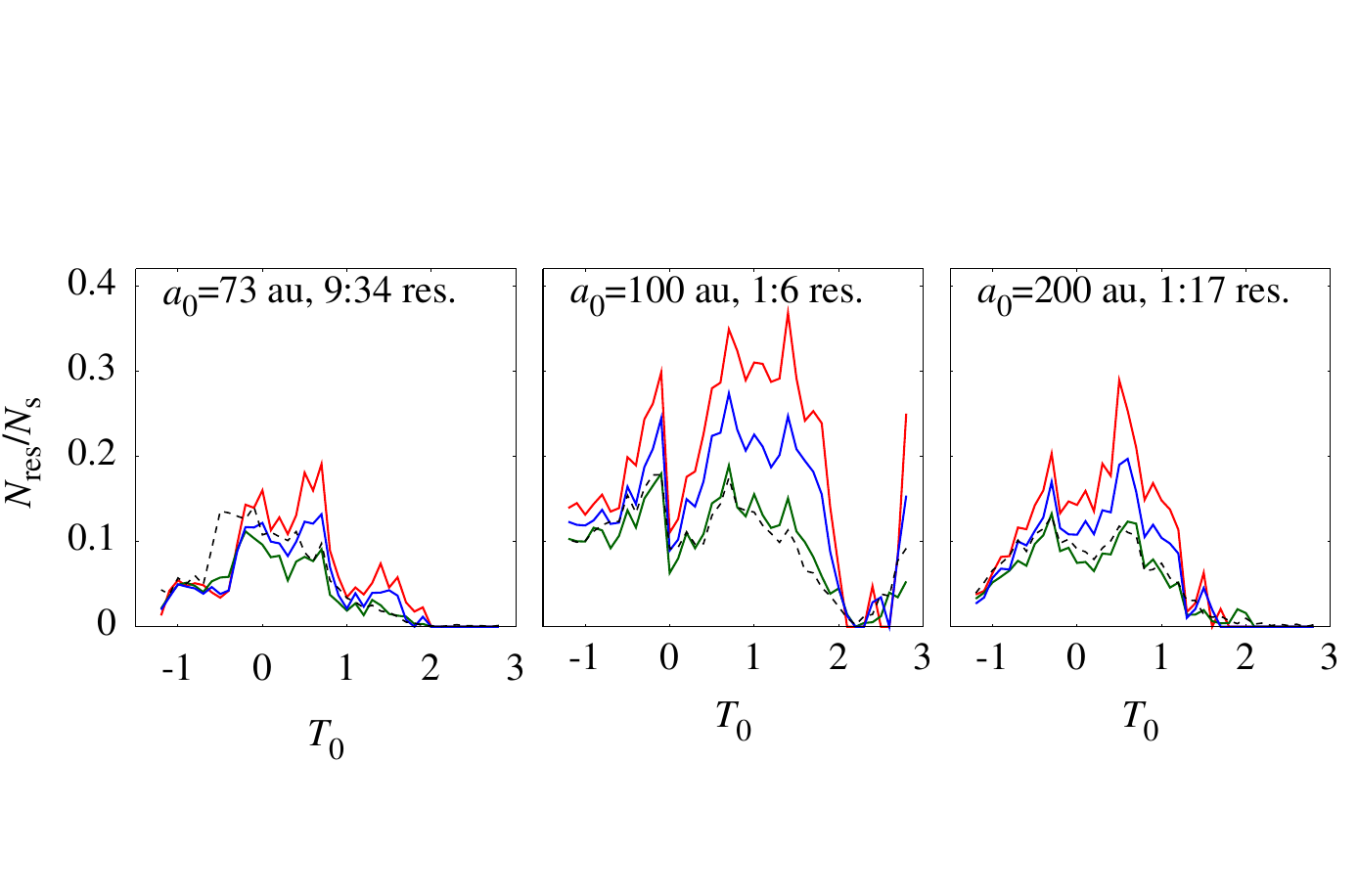} \label{f8}\vspace*{-15mm}
}
\begin{center}
\caption{Ratio of the resonant stable orbits, $N_{\rm res}$, to the total number of stable orbits $N_s$ at epoch $t$ as a function of the Tisserand invariant. The epochs are $t=0.1$\,Gyr (dashed black), 1\,Gyr (solid green), 3\,Gyr (solid blue) and $4.5$\,Gyr (solid red). Relevant resonances are shown next to the initial semi-major axis $a_0$. }
\end{center}\label{f8}
\end{figure}

\begin{figure}
{ 
\includegraphics[width=85mm]{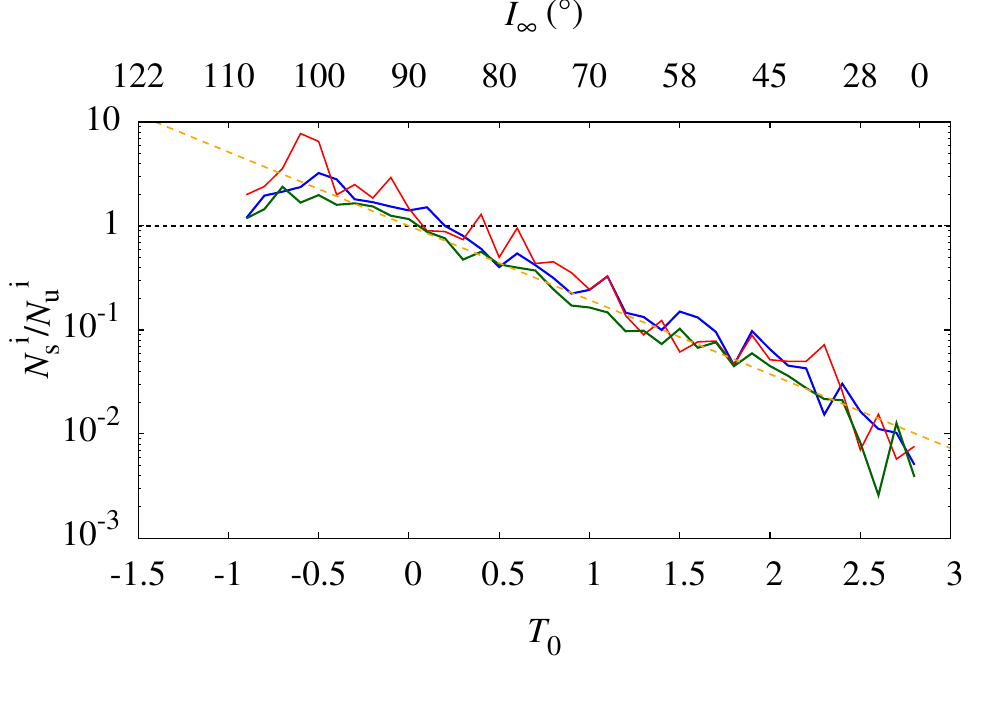} \label{f9}\vspace*{-3mm}
}
\begin{center}
\caption{Ratio of the number of injected stable orbits present at $4.5$\,Gyr, $N^{\rm i}_{\rm s}$, to the number of all injected unstable orbits from the simulation's start up to 4.5\,Gyr, $N^{\rm i}_{\rm u}$, as a function of the initial Tisserand invariant (and $I_\infty$) . The solid curves correspond respectively to $a_0=200$\,au (red), $100$\,au (blue) and $73$\,au (green).  The dashed line is $\log N^{\rm i}_{\rm s}/ N^{\rm i}_{\rm u}=-T_0/0.61$.}
\end{center}\label{f9}
\end{figure}

\begin{figure*}
{ 
\hspace*{-20mm}\includegraphics[width=110mm]{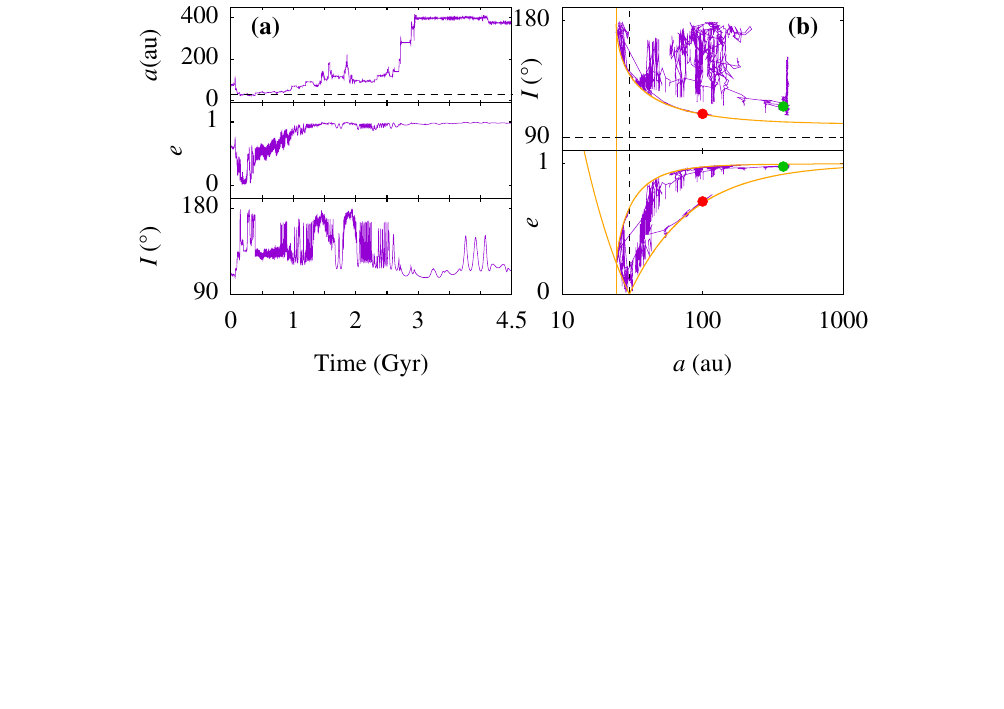}\hspace{-30mm}\includegraphics[width=110mm]{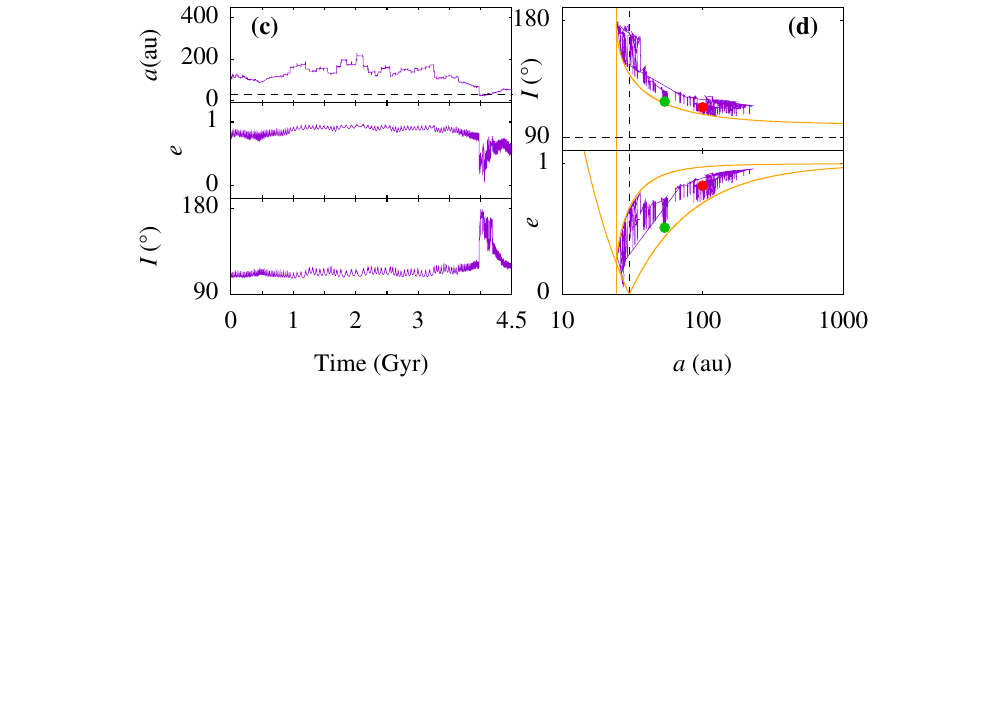}
\hspace{10mm} \label{f10}\vspace*{-30mm}
}
\begin{center}
\caption{Examples of collision singularity TNOs in the stable region of $a_0=100$\,au. The two TNOs have initial parameters $e_0=0.7112$, $I_0=108^\circ$, $\omega_0=154^\circ$  (panels a, b), and  $e_0=0.8313$, $I_0=113^\circ$, $\omega_0=92^\circ$  (panels c, d). Time evolution over $4.5$\, Gyr is shown in the left-hand panels of each asteroid (a, c) sampled every 1\,Myr. Neptune's position is indicated by the dashed black line in the semi-major axis panels. The TNOs' pathways in $(a,I)$ and $(a,e)$ planes are shown in the right-hand panels (b, d). The red and green full circles are the initial and final positions respectively. Also shown in (b, d) are the inclination pathway (\ref{TissIncP}), the perihelion and aphelion  crossing conditions, the maximal eccentricity (\ref{emax}), the planet's position (vertical dashed line) and the reflection semi-major axis (\ref{aX}) (orange vertical line). }
\end{center}\label{f10}
\end{figure*}

\begin{figure}
{ 
\includegraphics[width=95mm]{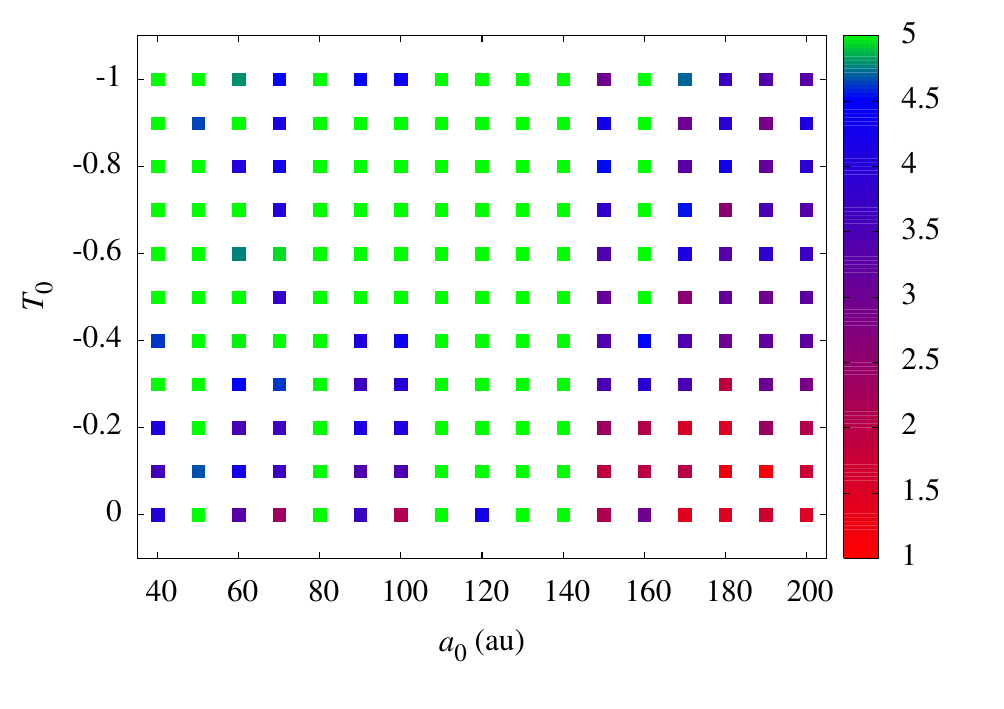} \label{f11}\vspace*{-3mm}
}
\begin{center}
\caption{Dynamical time portrait as a function of the initial semi-major axis $a_0$ and  Tisserand invariant, $T_0$. Color codes indicate the value of the dynamical time, $\tau_{\rm d}$ in Gyr. The simulation timespan is $5\,$Gyr implying that $\tau_{\rm d}$ at the green positions can exceed 5\,Gyr as shown for instance in Fig. 7.}
\end{center}\label{f11}
\end{figure}

\begin{figure}
{ 
\includegraphics[width=95mm]{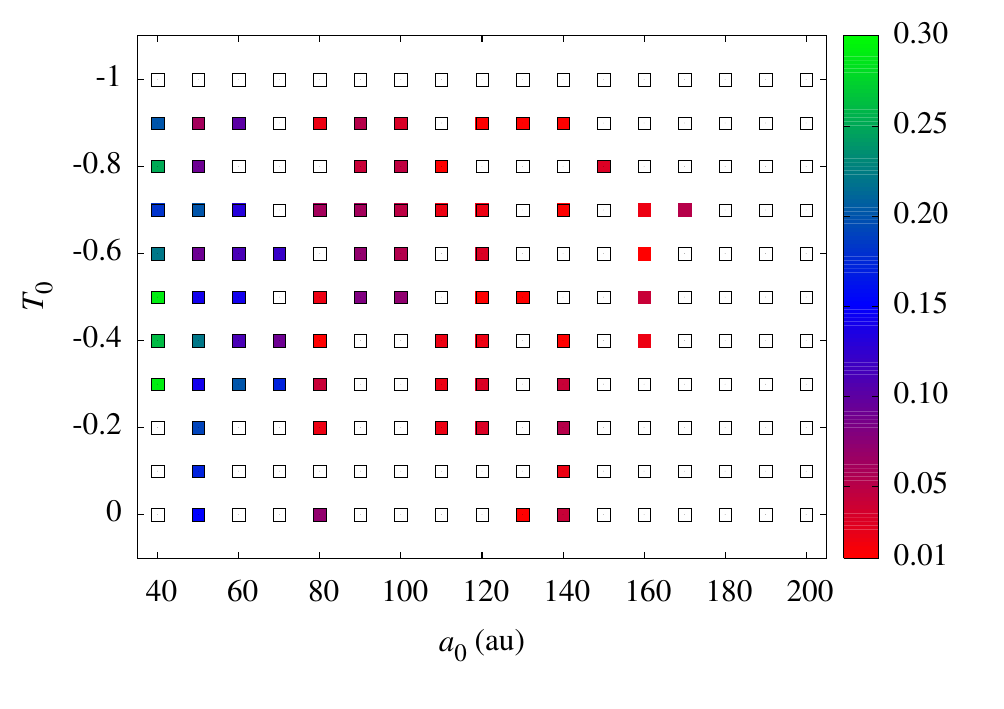} \label{f12}\vspace*{-3mm}
}
\begin{center}
\caption{Ratio of the number of injected stable orbits at  4.5\,Gyr, $N^{\rm i}_{\rm s}$ to the initial asteroid number as a function of the initial semi-major axis $a_0$ and  Tisserand invariant, $T_0$, for those locations with dynamical  times $\tau_{\rm d}\geq 4.5$\,Gyr and $N^{\rm i}_{\rm s}>0$. }
\end{center}\label{f12}
\end{figure}

\end{document}